# Inhomogeneity Generated Waves: Harbingers for Turbulence Generation

**Edisher Kh. Kaghashvili**

Ekathon Inc.

ekaghash@hotmail.com

**The Navier-Stokes equations describe fluid flow in many everyday life situations. Newton's second law of motion describes changes in the object's speed when a force applied. The Navier-Stokes equations are equivalent to Newton's Law when many objects such as microscopic particles in the air or water are considered as a group. Ocean currents, motion of smoke, dynamics of hurricane or weather patterns, air/fluid flow when the detonated device goes and many other interesting phenomena are all examples of situations when the Navier-Stokes equations govern temporal evolution. Despite its long history as a mathematical problem and simplicity of the equations, the time evolution of their solutions is not obvious. In this letter, I describe waves discovered some time ago that have received little attention in the neutral fluid physics community. Because of their properties these waves play an important role in processes like energy transfer, turbulence generation, heating, etc. Here, my intention is to demonstrate an existence of inhomogeneity generated waves in Navier-Stokes equations in the simplest possible case, give analytical solutions and provide some insight about them. We show (a) an existence of this new type of waves, which in our simple case considered here have amplitudes that increase with time, and (b) change in volumetric viscous heating when there is a shear flow present. Later process and accumulation of the kinetic energy in inhomogeneity generated waves shows an early stage of wave steepening in the linear regime.**

In 2002, while doing my thesis about the wave process in solar plasma, I came across a very interesting phenomenon: when an initial (incident) wave interacts with an inhomogeneity driving waves that were not known or described in the literature. I considered a specific example related to inhomogeneous flow with constant shear. In that particular case, I could not only show new types of waves numerically, but I also characterized them spatially and temporarily, and obtained analytical solutions. An unusual thing occurred in a simulation: one type of wave, specifically Alfvén wave named after its discoverer Hannes Alfvén (Ref. 1) generated waves of different frequencies when no mode coupling was expected according to existing studies of phase mixing and mode coupling. Inhomogeneity generated waves were originally shown in MagnitoHydroDynamics (**MHD)** when electro-magnetic effects are added to a system governed by Navier-Stokes equations). These were generalized for any physical system (Ref. 2-12). In what follows below, we will project our discovered inhomogeneity generated waves to a simple hydrodynamical case to show the novelty it brings in a picture.

Let us consider Navier-Stokes equations without dissipation. For simplicity we assume: (a) isothermal process (temperature remains constant, i.e., $p = c_s^2 \rho$ where $c_s^2 = const.$), (b) spatial variations are only in two-dimensional space (variations along z-axis are ignored), and (c) a simple one-dimensional Couette background flow. This simplification makes it easy to compare inhomogeneity generated waves with existing knowledge about wave processes in hydrodynamics. The governing equations read:

$$\frac{\partial \rho}{\partial t} + \nabla(\rho \boldsymbol{u}) = 0 \tag{1}$$

$$\rho \left[\frac{\partial \boldsymbol{u}}{\partial t} + (\boldsymbol{u} \cdot \nabla)\boldsymbol{u}\right] = -c_s^2 \nabla \rho \tag{2}$$

where $\rho$ is the density of the fluid and $\boldsymbol{u}$ is the two-dimensional flow. For simplicity, we will assume only background inhomogeneity is present in the fluid, namely inhomogeneous flow directed along the x-axis and shared along the y-axis: $\boldsymbol{U_0} = (U_0, 0, 0)$, where $U_0 = S_y y$ and $S_y = const.$

Linear waves in this system are governed by

$$\frac{\partial u_x}{\partial t} + ik_x c_s^2 D = -S_y u_y \tag{3}$$

$$\frac{\partial u_y}{\partial t} + iK_y c_s^2 D = 0 \tag{4}$$

Together with the mass conservation equation

$$\frac{\partial D}{\partial t} + ik_x u_x + iK_y u_y = 0 \tag{5}$$

where a linear wave is represented as $f \Rightarrow f \exp(ik_x x + iK_y y)$. Other variables in (3)-(5) are as follows: $D = \rho'/\rho_0$, $k_x = const.$, $K_y = k_y - k_x S_y t$, and $k_y = K_y(t = 0) = const.$

Let us consider one specific initial condition to solve the above equations that leads to a solution which can be easily looked up in any book on the subject, namely an evolution of initial perturbation when only $D(t = 0) = D_0$ is present at $t = 0$. When there is no shear flow present (i.e., $S_y = 0$), above equations can be readily solved:

$$D(t) = D_0 \cos(\omega_{s0} t) \tag{6}$$

$$u_x(t) = -\frac{ik_x D_0 c_s}{k} \sin(\omega_{s0} t) \tag{7}$$

$$u_y(t) = -\frac{ik_y D_0 c_s}{k} \sin(\omega_{s0} t) \tag{8}$$

where $\omega_{s0} = c_s \sqrt{k_x^2 + k_y^2}$ is the oscillation frequency of a sound wave, and $k = \sqrt{k_x^2 + k_y^2}$. Figure 1a and 1b show the result of the simulations of (3)-(5) system. Solutions given by (6)-(8) exactly match simulation results shown on the left-hand side column (Figure 1a; this is when there is no shear flow: $S_y = 0$). These sound wave solutions can be found in the literature about the subject. On the other hand, Figure 1b (the right-hand side column of Figure 1) presents the solutions with inhomogeneous flow included.

**Are time series in Figure 1a and Figure 1b different?** At later time in Figure 1b, one can notice that there is an energy exchange between density and momentum fluctuations in case of inhomogeneous flow even in a linear regime something that is not present in Figure 1a. What about other differences? Let us examine a difference in time series. The result is shown in Figure 2. As we can see the changes in all fluctuating variables are significant. Next, we will investigate the origin of these differences in fluctuations.

Following an inhomogeneity generated wave formalism (see, e.g., Ref. 2-3), we can represent the solutions of (3)-(5) as a series in terms of the shear parameter:

$$D(t) = D_{s0}(t) + S_y D_{s1}(t) + S_y^2 D_{s2}(t) + \ldots \tag{9}$$

$$u_x(t) = u_{x0}(t) + S_y u_{x1}(t) + S_y^2 u_{x2}(t) + \ldots \tag{10}$$

$$u_y(t) = u_{y0}(t) + S_y u_{y1}(t) + S_y^2 u_{y2}(t) + \ldots \tag{11}$$

where zero-order shear parameter terms are often given in the literature, and **all other subsequent terms are explicitly $S_y$-parameter dependent and represent inhomogeneity generated waves**. We first derive second order wave equations for the fluctuating variables, and substitute above expansions in the second order governing equations. Afterwards, we group those equations according to their order in terms of the shear parameter, $S_y$.

Most of the applications of the inhomogeneity generated waves I considered previously (Ref. 4-12) were about processes in the magnetized plasma where initial wave was chosen to be the most stable mode in that system, namely Alfvén wave with frequency that was not changing with time (and it's amplitude too). Here our initial disturbance will propagate with gradually varying wavelength and amplitude. As was shown above, there is only one characteristic frequency in our system, and it changes with time as

$$\omega_s(t) = c_s \sqrt{k_x^2 + K_y^2(t)}. \tag{12}$$

Hence, we need to use a well-known WKB approximation (Ref. 13) when solving for zero-order solutions. In this approximation, it is assumed that the medium varies gradually relative to the wavelength. At zero-order, the solutions are:

$$D_{s0}(t) = D_0 \sqrt{\frac{k_x^2 + k_y^2}{k_x^2 + K_y^2(t)}} \cos\left(\int_0^t \omega_s(\tau) d\tau\right) \equiv A_D(t) \cos\left(\int_0^t \omega_s(\tau) d\tau\right) \tag{13}$$

$$u_{x0}(t) = -i k_x c_s^2 D_0 \int_0^t \left(\sqrt{\frac{k_x^2 + k_y^2}{k_x^2 + K_y^2(\tau)}} \cos\left(\int_0^\tau \omega_s(\rho) d\rho\right)\right) d\tau \tag{14}$$

$$u_{y0}(t) = -ic_s^2 D_0 \int_0^t \left( K_y(\tau) \sqrt{\frac{k_x^2+k_y^2}{k_x^2+K_y^2(\tau)}} \cos\left(\int_0^\tau \omega_s(\rho)d\rho\right) \right) d\tau \qquad (15)$$

where $A_D(t)$ is the amplitude of the density fluctuations. Deriving these zero-order solutions, WKB approximation assuming that perturbations evolve adiabatically, which in mathematical terms translates as a condition on changes in their amplitudes. As an example, for density in (13), we have the condition: $\ddot{A}_D(t) \ll A_D(t)$. Figure 3a compares the zero-order WKB solutions given by (13)-(15) with numerical simulations. As can been seen WKB solutions closely describe what is happening in the system. Nevertheless, there are changes in the waveform that WKB does not capture. Figure 3b shows the deference between numerical and WKB-analytical solutions, these differences seem to be waves as well with time increasing amplitudes.

**So, what additional waves are generated in the system?**

It is straightforward to verify that neglected terms in WKB solutions are second order in terms of the small shear-parameter, and no such amplitude increase expected for them. Quite an opposite, these neglected terms involving the second derivative of the wave amplitudes vanish rapidly with time. Can inhomogeneity generated waves introduced in (9)-(11) explain these additional waves in the system? To answer this question, let us solve for the first-order, $S_y$-parameter proportional terms in the equations.

A governing equation for the first order density term is:

$$\frac{\partial^2 D_{s1}}{\partial t^2} + c_s^2 \left(k_x^2 + K_y^2(t)\right) D_{s1} = 2ik_x u_{y0} \qquad (16)$$

where initial conditions for the first-order inhomogeneity generate density wave are: $D_{s1}(t=0) = 0$, and $\partial D_{s1}/\partial t|_{t=0} = 0$. Equation (16) states what is postulated in the inhomogeneity generated wave formalism (e.g., Ref. 2, 10), namely, zero-order quantities (initial waves) drive the first order inhomogeneity generated waves, and first-order inhomogeneity generated waves drive the second order inhomogeneity generated waves and so on. **Since there is only one**

**natural frequency in the system and the driver term also oscillates with the same frequency, the resonance effect is expected to produce time dependent amplitude inhomogeneity generated waves.**

Figure 4 shows the result of the simulation of equation (16) that is compared to a difference between actual inhomogeneous flow simulation and WKB approximation solution shown in Figure 3b. As can be seen, there is a qualitative and quantitative match. We leave it up to the reader to confirm that a difference between these two time series is also a wave corresponding to the second-order inhomogeneity generated waves and their amplitudes also increase with time. Similar kind analysis can be done for fluctuating velocity components in Figure 3b.

Another important aspect of the process is the wave heating. Figure 5 shows a volume average viscous dissipation of these waves in all three cases: (a) no shear flow, (b) WKB case, and (c) shear flow present ($S_y = 0.1$). As can be seen, there is a less dissipation in the simulations when there is a shear flow present. Less heating and accumulation of the kinetic energy in inhomogeneity generated waves as shown in Figure 3b would lead to shock formation. Hence excitement of the inhomogeneity generated waves in the system shows an early stage of this commonly regarded non-linear phenomenon in the linear regime.

**To summarize**, we considered a simple physical process when only inhomogeneous flow was present in the system. In a non-dissipative hydro dynamical system governed by the Navier-Stokes equations, inhomogeneity generated waves can be generated by flow, density, pressure or temperature spatial dependence as shown below by underlined terms:

$$\frac{\partial \rho}{\partial t} + (\boldsymbol{U}_0 \cdot \nabla)\rho + \underline{\rho(\nabla \cdot \boldsymbol{U}_0)} + \rho_0(\nabla \cdot \boldsymbol{u}) + \underline{(\boldsymbol{u} \cdot \nabla)\rho_0} = 0 \tag{17}$$

$$\frac{\partial \boldsymbol{u}}{\partial t} + (\boldsymbol{U}_0 \cdot \nabla)\boldsymbol{u} + \underline{(\boldsymbol{u} \cdot \nabla)\boldsymbol{U}_0} - \underline{\frac{\rho}{\rho_0^2}\nabla p_0} = -\nabla p \tag{18}$$

$$\frac{\partial p}{\partial t} + (\boldsymbol{U}_0 \cdot \nabla)p + \underline{(\boldsymbol{u} \cdot \nabla)p_0} + \gamma p_0(\nabla \cdot \boldsymbol{u}) + \underline{\gamma p(\nabla \cdot \boldsymbol{U}_0)} = 0 \tag{19}$$

where pressure is a product of temperature and density. Any initial disturbance propagating in this system is expected to be coupled with the inhomogeneity rates in underlined terms and excite linear inhomogeneity generated waves.

The processes in nature are more complicated. The references below (Ref. 2-12) provide more detailed information about (a) inhomogeneity generated waves, (b) the formalism introduced to obtain analytical solutions, and (c) the application of these waves in many observed phenomena in solar, magnetosphere, and laboratory physics. Here I present some basic properties of inhomogeneity generated waves independent of the physical system (Hydrodynamics, MHD, electron-proton plasma, etc.) derived from our earlier works:

a) **Type of Waves.** These new waves are generated linearly when any initial disturbance (wave, spike, pulse, etc.) interacts with any background inhomogeneity (in background flow, density, pressure, temperature, etc.),

b) **Basic Features.** Some essential features of inhomogeneity generated waves are:
   - Wave amplitudes are proportional to inhomogeneity rates (i.e., density gradients, flow inhomogeneity, sharing of the magnetic field, etc.),
   - Expected natural frequencies of these waves also are modified, and their polarization properties are often different than what are expected from existing knowledge,
   - Generated waves can become the source of new waves, and
   - The most interesting case is when a stable wave of the system propagates in the inhomogeneous medium (i.e., the wave energy is transferred over long distances). It continuously generates inhomogeneity-rate-dependent-amplitude waves during this process. Generated waves tend to be more dissipative thus effectively distributing energy in the system at different scales.

c) **How many waves per any given initial disturbance**? In general, any initial disturbance will generate waves with all possible natural frequencies of the system that governs the phenomenon.

d) **Heating/Dissipation**. When electro-magnetic effects are included, inhomogeneity generated waves can generate a strong electric field that can interact with particles. In Hydrodynamics, it is also expected that inhomogeneity generated waves shown in Figure

3b will dissipate more effectively than initial sound waves.

e) **Time Evolution.** Since the analytical description shows only what happens immediately after an initial disturbance interacts with inhomogeneity, the time evolution of the waves/medium/flow/plasma is expected to be a more complex process involving (a) continues withdrawal of available energy by inhomogeneity generated waves and its transport, and (b) continues breaking the existing or just generated waves due to dissipation. This leads to effective spatial and temporal energy distribution in the system.

**Acknowledgements:** I would like to thank Frank Frazier for his assistance in the review and editing of the manuscript.

**REFERENCES**

1. Hannes Alfven, https://en.wikipedia.org/wiki/Hannes_Alfvén

2. Kaghashvili E. Kh., 2014, High-frequency driven waves in the solar atmosphere, ArXiv

3. Kaghashvili E. Kh., 2013, Alfvén waves in shear flows: Driven wave formalism, *J. Plasma Physics*, 79, 797, doi:10.1017/S0022377813000500

4. Hollweg J. V., Kaghashvili E. Kh. And B. D.G. Chandran, 2013, Velocity-shear-induced mode coupling in the solar atmosphere and solar wind: Implications for plasma heating and MHD turbulence, *Astrophys. J.*, **769**, 142

5. Kaghashvili E., 2012, Driven wave generated electric field in the solar corona, *J. of Geophys. Res.*, **117**, A10103, doi:10.1029/2012JA018120

6. Kaghashvili E. Kh., 2012, Alfvén wave dynamics in the expanding solar corona, edited by T. R. Rimmele, A. Tritschler, F. Woger, et al., *Astron. Soc. Pacific Conf. Ser.*, **463**, 175

7. Kaghashvili E. Kh., 2012, Beyond standard coronal seismology tools, edited by T. R. Rimmele, A. Tritschler, F. Woger, et al., *Astron. Soc. Pacific Conf. Ser.*, **463**, 235

8. Hollweg J. V. and E. Kh. Kaghashvili, 2012, Alfvén waves in shear flows revisited, *Astrophys. J.*, **744**, 114

9. Kaghashvili E.Kh., R. A. Quinn and J. V. Hollweg, 2009, Driven waves as a diagnostics tool in the solar corona, *Astrophys. J.*, **703**, 1318

10. Kaghashvili E. Kh., Raeder J., Webb, G. M. and G. P. Zank, 2007, Propagation of Alfvén waves in shear flows: perturbation method for driven fluctuations, edited by D. Shaikh and G. P. Zank, *American Institute of PhysicsConf. Proc.*, **932**, 415

11. Kaghashvili E. Kh., 2007, Alfvén wave driven compressional fluctuations in shear flows, *Physics of Plasmas*, **14**, 044502

12. Kaghashvili E. Kh., J. Raeder, Webb G. M. and G. P. Zank, 2006, Propagation of Alfvén waves in shear flows: nature of driven longitudinal velocity and density fluctuations, *Physics of Plasma*, **13**, 112107

13. WKB Approximation: https://en.wikipedia.org/wiki/WKB_approximation

**FIGURE 1**: Time evolution of the density and velocity components. Figure 1a (left-hand side column) corresponds to a case with no inhomogeneous flow. Figure 1b (right-hand side column) is a case with inhomogeneous flow. In both cases, the blue line represents the real part of the corresponding variable's temporal evolution. The red line denotes the imaginary component. Normalization parameters are $\tau = tk_x c_s$, $u_{\{x,y\}} \Rightarrow u_{\{x,y\}}/c_s$, $k_y/k_x = 8$, $S_y = 0.1$, $u_x(t=0)=0$, $u_y(t=0)=0$, and $D(t=0) = D_0 = 0.1$.

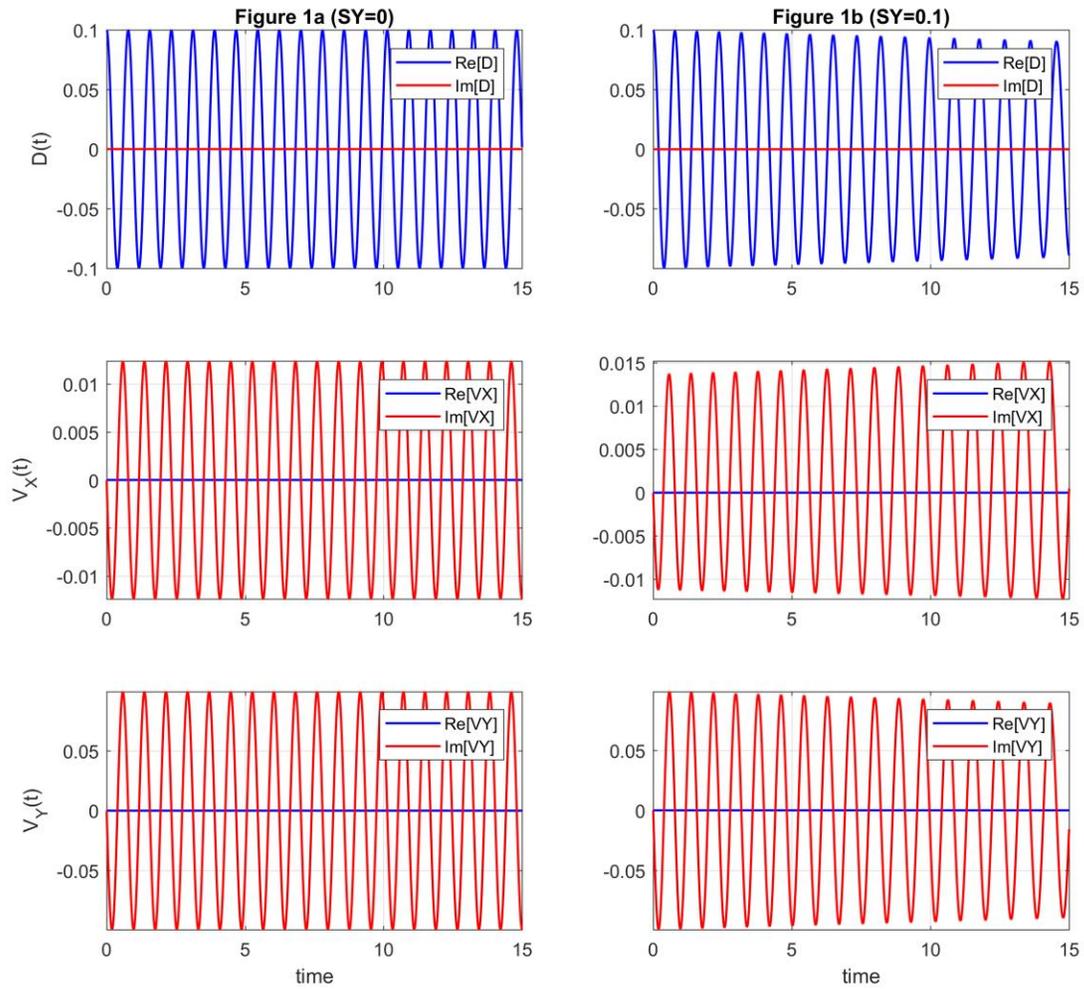

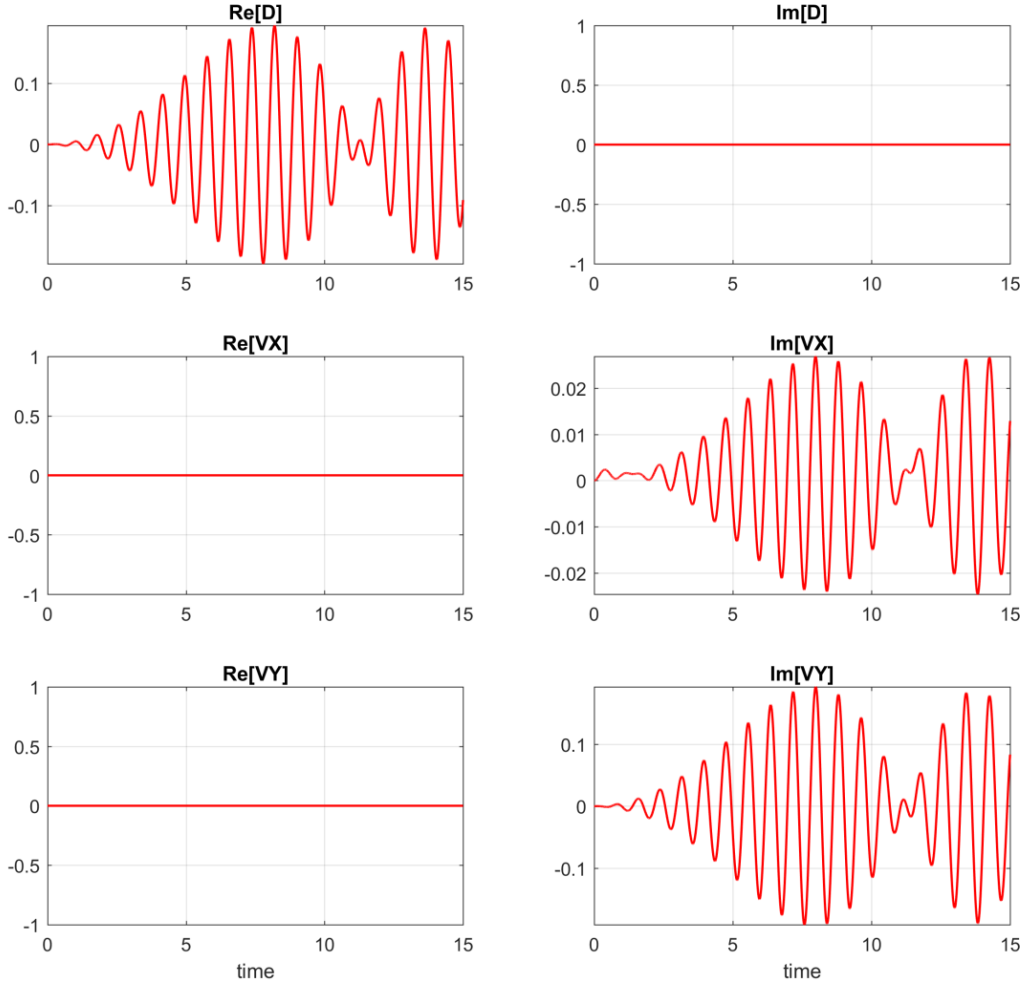

**FIGURE 2**: Difference between two cases presented in Figure 1a and Figure 1b.

**FIGURE 3**: Fluctuating components in two cases: (a) simulations when $S_y = 0.1$ (blue line in Figure3a), and (b) WKB solutions given by the equations (13)-(15) above (red dashed line in Figure 3a). Figure 3a (left-hand side column) provides both time series. It shows that WKB solution accounts for the frequency shift. Figure 3b (right-hand side column) shows differences between these two time series.

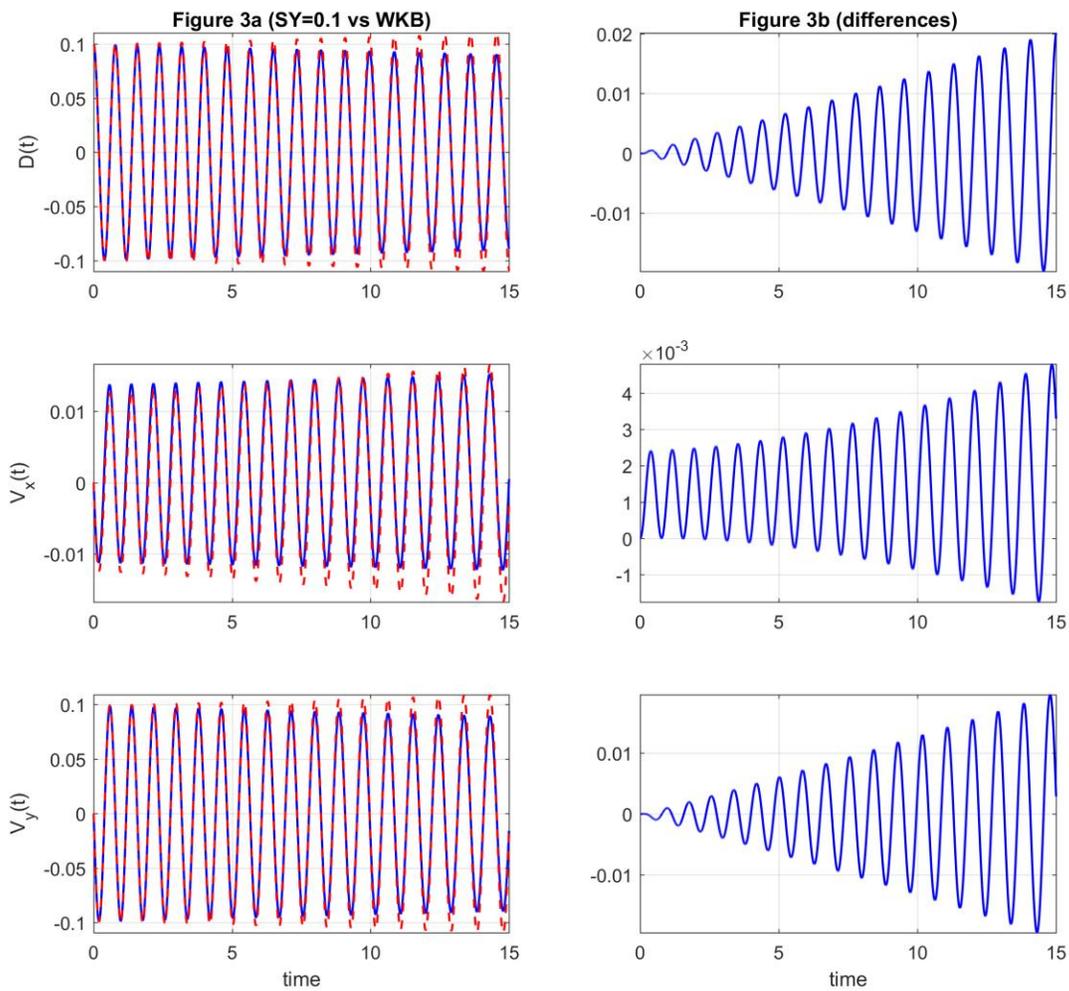

**FIGURE 4**: Real density components in two cases: (a) shown in Figure 3b (blue line), which is a difference between the inhomogeneous flow simulations minus the WKB solution, and (b) the first-order inhomogeneity generated wave density solution.

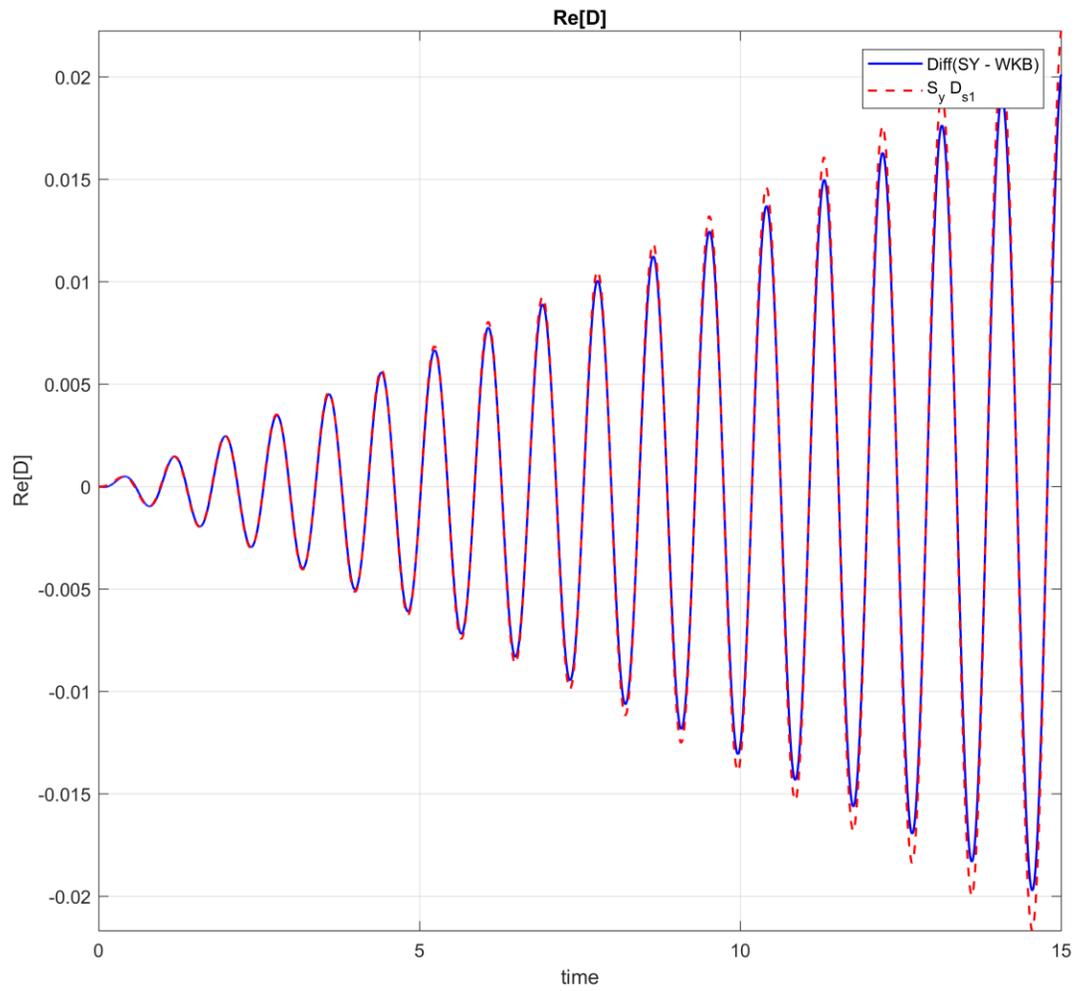

**FIGURE 5**: Volume average viscous dissipation of inhomogeneity generated waves in all three cases: (a) no shear flow ($S_y = 0$), (b) WKB case, and (c) shear flow present ($S_y = 0.1$). All volumetric heating terms are normalized to total dissipated energy value in case of no shear flow. As shown, in this particular case, there is a least volumetric viscous dissipation in the simulation when there is a shear flow present.

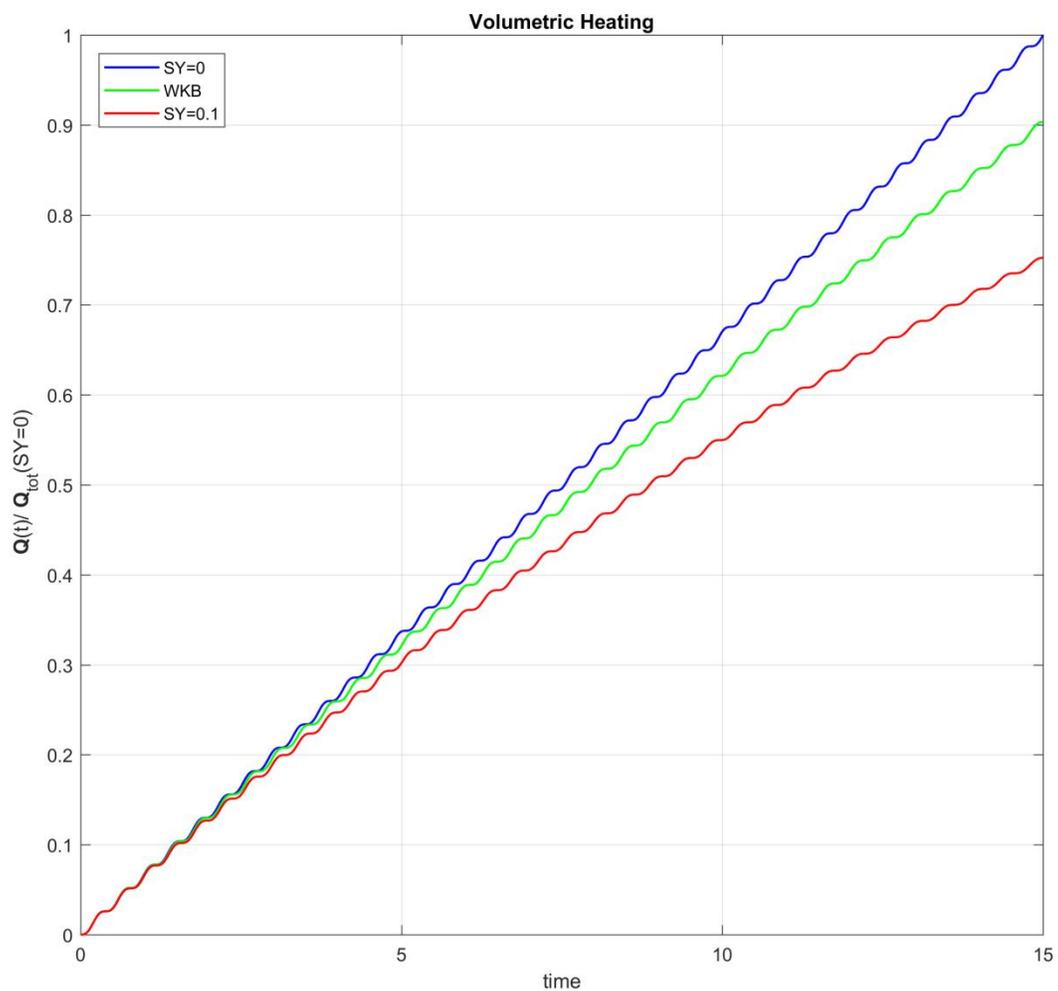